\documentclass[epj]{svjour}
\usepackage{amsmath,amssymb,feynarts}
\usepackage{colordvi,gnuplotcolors}

\usepackage{graphicx}
\usepackage{fancyhdr}
\def\makeheadbox{{
 }}
\def\mytitle{My title} 
\def\myauthors{My name}  
\def\mytype{My type of session}
\def\mysession{My session}

\def\mytitle{Distributions for Higgs + jet at hadron colliders: MSSM versus SM} 
\def\myauthors{Oliver Brein, Wolfgang Hollik}    
\def\mytype{Contributed Talk}    
\def\mysession{Colliders - Higgs Phenomenology}

\newcommand{\gev}{\text{GeV}}
\newcommand{\tev}{\text{TeV}}
\newcommand{\fb}{\text{fb}}
\newcommand{\pb}{\text{pb}}

\newcommand{\ma}{ m_{A^0} }

\newcommand{\MSUSY}{ M_{\text{SUSY}} }

\newcommand{\MAX}{{\text{max}}}

\newcommand{\JET}{\text{jet}}

\newcommand{\MSSM}{{\text{MSSM}}}
\newcommand{\SM}{{\text{SM}}}

\newcommand{\tb}{\tan\beta}

\begin{document}
\title{Distributions for Higgs + jet at hadron colliders:\\
 MSSM vs SM}
\subtitle{}
\author{Oliver Brein\inst{1}
\thanks{\emph{Email:} Oliver.Brein@durham.ac.uk (speaker)}%
 \and
 Wolfgang Hollik\inst{2}
\thanks{\emph{Email:} hollik@mppmu.mpg.de}%
}                     
%
%
\institute{Institute for Particle Physics Phenomenology,
University of Durham, DH1 3LE, Durham, United Kingdom
\and Max-Planck-Institut f\"ur Physik,
        F\"ohringer Ring 6, D-80805 M\"unchen, Germany
}
%
\date{}
\abstract{
We present pseudorapidity and transverse momentum distributions for
the cross section for the production of the lightest
neutral Higgs boson in association
with a high-$p_T$ hadronic jet, calculated in the framework of
the minimal supersymmetric standard model.
We discuss the theoretical predictions for the differential
cross sections at the Large Hadron Collider and the Tevatron.
In particular, we present the differences in the distributions
compared to the Standard Model.
\PACS{
      {PACS-key}{discribing text of that key}   \and
      {PACS-key}{discribing text of that key}
     } 
} 
\maketitle
%

\section{Higgs + Jet in the Standard Model}
\label{sec:hjet-in-SM}

The production of SM Higgs bosons in hadron collisions at the 
LHC will proceed mainly via gluon fusion ($gg\to H$).
The detection of a SM Higgs boson
with a mass below 130$\,\gev$
at the LHC is rather difficult
because the predominant decay
into a $b\bar b$-pair is swamped by the large QCD two-jet 
background \cite{CMS-TDR}.
Therefore, only through observation of the
rare decay into two photons
is the inclusive single Higgs boson production 
considered the best search channel 
in this mass range at the LHC.

Alternatively,  
and in order to fully explore the Higgs-detection capabilities 
of the LHC detectors, 
one can investigate more exclusive channels 
like e.g.~Higgs production in association with a high-$p_T$ hadronic jet
\cite{hjet-sm}.
The main advantage of this channel is the richer kinematical structure 
of the events
which 
allows for refined cuts increasing the 
signal-to-background 
ratio, obtained at the price of a 
lower 
signal rate compared to
the inclusive channel (about $10\%$
of the rate of the inclusive process).
For the SM Higgs boson,
simulations of this signal process and its background,
considering the decay channels
$H\rightarrow \gamma\gamma$ \cite{ADIKSS,Zmushko} 
and $H\rightarrow \tau^+ \tau^-$ \cite{mellado-etal},
have shown promising results for the ATLAS detector.
Also, very recently, promising simulation results for the Tevatron
appeared involving this process in connection with the decay $H\to W^+W^-$
\cite{new-mellado-etal}.

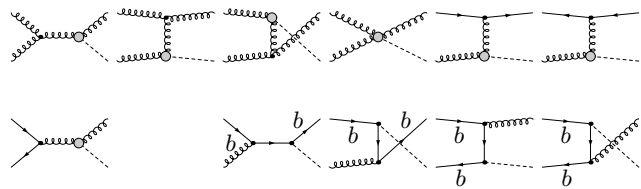
\begin{figure}[bt]
\begin{footnotesize}

\begin{picture}(378,80)(20,0)
\unitlength=1bp%
\begin{feynartspicture}(270,80)(6,2)
\FADiagram{}
\FAProp(0.,15.)(6.,10.)(0.,){/Cycles}{0}
\FAProp(0.,5.)(6.,10.)(0.,){/Cycles}{0}
\FAProp(20.,15.)(14.,10.)(0.,){/Cycles}{0}
\FAProp(20.,5.)(14.,10.)(0.,){/ScalarDash}{0}
\FAProp(6.,10.)(14.,10.)(0.,){/Cycles}{0}
\FAVert(6.,10.){0}
\FAVert(14.,10.){-1}

\FADiagram{}
\FAProp(0.,15.)(10.,14.)(0.,){/Cycles}{0}
\FAProp(0.,5.)(10.,6.)(0.,){/Cycles}{0}
\FAProp(20.,15.)(10.,14.)(0.,){/Cycles}{0}
\FAProp(20.,5.)(10.,6.)(0.,){/ScalarDash}{0}
\FAProp(10.,14.)(10.,6.)(0.,){/Cycles}{0}
\FAVert(10.,14.){0}
\FAVert(10.,6.){-1}

\FADiagram{}
\FAProp(0.,15.)(10.,14.)(0.,){/Cycles}{0}
\FAProp(0.,5.)(10.,6.)(0.,){/Cycles}{0}
\FAProp(20.,15.)(10.,6.)(0.,){/Cycles}{0}
\FAProp(20.,5.)(10.,14.)(0.,){/ScalarDash}{0}
\FAProp(10.,14.)(10.,6.)(0.,){/Cycles}{0}
\FAVert(10.,6.){0}
\FAVert(10.,14.){-1}

\FADiagram{}
\FAProp(0.,15.)(10.,10.)(0.,){/Cycles}{0}
\FAProp(0.,5.)(10.,10.)(0.,){/Cycles}{0}
\FAProp(20.,15.)(10.,10.)(0.,){/Cycles}{0}
\FAProp(20.,5.)(10.,10.)(0.,){/ScalarDash}{0}
\FAVert(10.,10.){-1}

\FADiagram{}
\FAProp(0.,15.)(10.,14.)(0.,){/Straight}{1}
\FAProp(0.,5.)(10.,6.)(0.,){/Cycles}{0}
\FAProp(20.,15.)(10.,14.)(0.,){/Straight}{-1}
\FAProp(20.,5.)(10.,6.)(0.,){/ScalarDash}{0}
\FAProp(10.,14.)(10.,6.)(0.,){/Cycles}{0}
\FAVert(10.,14.){0}
\FAVert(10.,6.){-1}

\FADiagram{}
\FAProp(0.,15.)(10.,14.)(0.,){/Straight}{-1} 
\FAProp(0.,5.)(10.,6.)(0.,){/Cycles}{0}
\FAProp(20.,15.)(10.,14.)(0.,){/Straight}{1}
\FAProp(20.,5.)(10.,6.)(0.,){/ScalarDash}{0}
\FAProp(10.,14.)(10.,6.)(0.,){/Cycles}{0}
\FAVert(10.,14.){0}
\FAVert(10.,6.){-1}

\FADiagram{}
\FAProp(0.,15.)(6.,10.)(0.,){/Straight}{1}
\FAProp(0.,5.)(6.,10.)(0.,){/Straight}{-1}
\FAProp(20.,15.)(14.,10.)(0.,){/Cycles}{0}
\FAProp(20.,5.)(14.,10.)(0.,){/ScalarDash}{0}
\FAProp(6.,10.)(14.,10.)(0.,){/Cycles}{0}
\FAVert(6.,10.){0}
\FAVert(14.,10.){-1}

\FADiagram{}

\FADiagram{}
\FAProp(0.,15.)(6.,10.)(0.,){/Straight}{1}
\FALabel(2.48771,11.7893)[tr]{$b$}
\FAProp(0.,5.)(6.,10.)(0.,){/Cycles}{0}
\FAProp(20.,15.)(14.,10.)(0.,){/Straight}{-1}
\FALabel(16.4877,13.2107)[br]{$b$}
\FAProp(20.,5.)(14.,10.)(0.,){/ScalarDash}{0}
\FAProp(6.,10.)(14.,10.)(0.,){/Straight}{1}
\FAVert(6.,10.){0}
\FAVert(14.,10.){0}

\FADiagram{}
\FAProp(0.,15.)(10.,14.)(0.,){/Straight}{1}
\FALabel(4.84577,13.4377)[t]{$b$}
\FAProp(0.,5.)(10.,6.)(0.,){/Cycles}{0}
\FAProp(20.,15.)(10.,6.)(0.,){/Straight}{-1}
\FALabel(16.8128,13.2058)[br]{$b$}
\FAProp(20.,5.)(10.,14.)(0.,){/ScalarDash}{0}
\FAProp(10.,14.)(10.,6.)(0.,){/Straight}{1}
\FAVert(10.,14.){0}
\FAVert(10.,6.){0}

\FADiagram{}
\FAProp(0.,15.)(10.,14.)(0.,){/Straight}{1}
\FALabel(4.84577,13.4377)[t]{$b$}
\FAProp(0.,5.)(10.,6.)(0.,){/Straight}{-1}
\FALabel(5.15423,4.43769)[t]{$b$}
\FAProp(20.,15.)(10.,14.)(0.,){/Cycles}{0}
\FAProp(20.,5.)(10.,6.)(0.,){/ScalarDash}{0}
\FAProp(10.,14.)(10.,6.)(0.,){/Straight}{1}
\FAVert(10.,14.){0}
\FAVert(10.,6.){0}

\FADiagram{}
\FAProp(0.,15.)(10.,14.)(0.,){/Straight}{1}
\FALabel(4.84577,13.4377)[t]{$b$}
\FAProp(0.,5.)(10.,6.)(0.,){/Straight}{-1}
\FALabel(5.15423,4.43769)[t]{$b$}
\FAProp(20.,15.)(10.,6.)(0.,){/Cycles}{0}
\FAProp(20.,5.)(10.,14.)(0.,){/ScalarDash}{0}
\FAProp(10.,14.)(10.,6.)(0.,){/Straight}{1}
\FAVert(10.,14.){0}
\FAVert(10.,6.){0}
\end{feynartspicture}
\end{picture}

\end{footnotesize}

\caption{\label{SM_parton_processes}
Partonic processes contributing to 
$pp \to H + \text{jet} + X$ in the SM.
Hatched circles represent
loops of heavy quarks. The depicted tree-level $b$-quark processes
are much more important in the MSSM case.}
\end{figure}

The partonic processes at leading order
contributing to the hadronic reaction 
$pp \to H + \text{jet} + X$ (see Fig.~\ref{SM_parton_processes}) 
are gluon fusion ($gg \to g H$, 50--70 \% of total rate), 
quark--gluon scattering 
($q(\bar q)g \to q(\bar q) H$, 30--50 \% of total rate)
and 
quark--antiquark annihilation ($q\bar q \to g H$, rate small).
The hadronic 
cross section is dominated by loop-induced processes, involving
effective $ggH$- and $ggHZ$-coup\-lings. If the $b$-quark is treated as a 
parton present in the proton, there are additional tree-level processes
for quark-gluon scattering and quark-antiquark annihilation
to consider. Yet in the SM, their contribution to the hadronic cross
section is small.

A lot of progress has been made
towards improving the SM predictions.
The fully differential distribution for Higgs production
at next-to-next-to-leading order QCD accuracy
has become available \cite{AMP}, improved by
resummation of logarithmically enhanced  
terms for low $p_T$~\cite{grazzini-etal}.
Higher-order corrections to differential cross sections for 
a Higgs boson associated with a high-$p_T$ jet
have been obtained explicitly: the next-to-leading order QCD corrections 
in the large top-mass limit \cite{kunszt-etal}
and, recently, the corresponding resummation of soft-gluon 
emission effects \cite{kulesza-etal}.
For the $b$-quark process $b g\to H b$, the NLO QCD corrections 
are also known \cite{bghb-QCD}.

\section{Higgs + Jet in the MSSM}
\label{sec:hjet-in-MSSM}

Motivated by the promising SM simulation\cite{ADIKSS,Zmushko}
we investigated the MSSM process $pp\to h^0+\text{jet}+X$,
involving the lightest MSSM Higgs boson \cite{hjet-own}.
Especially, as the process is essentially 
loop-induced, there are potentially large effects from virtual
superpartners to be expected.

In the meantime, for this process,
$p_T$-distributions 
have been studied 
in the limit of vanishing superpartner contributions 
at leading order~\cite{BField-etal} and were
improved recently by soft-gluon
resummation effects \cite{langenegger-etal}.
This limit is 
usually a good approximation when the superpartners
are heavy, at a mass scale around $1\,\tev$.
Quite recently, the SUSY-QCD corrections to the 
cross section and $p_T$ distribution of the $b$-quark
initiated processes $b g\to h^0 b$ have been calculated
\cite{dawson-jackson}.

In the MSSM, the classes of contributing partonic processes are 
basically the similar to the SM:
gluon fusion $g+g\to g+h^0$,
quark--gluon scattering $q+g\to q+h^0$,
and 
quark--anti-quark annihilation $q +\bar q \to g+ h^0$.
While gluon fusion is an entirely loop-induced process, 
the other two classes also 
get contributions from tree-level $b$-quark initiated processes 
(see Fig.~\ref{SM_parton_processes}).
Those Born-type processes are in general 
dominant for $m_A \lesssim 120\,\gev$,
while for large values of $m_A$ the loop-induced processes dominate. 
This behaviour is  essentially a consequence of the 
Yukawa-coupling of the lightest MSSM Higgs boson
to $b$-quarks, which can be enhanced for low values of the $A$-boson mass
$m_A$.
Formulae for the relevant couplings occurring in this process
in our notation can be found in \cite{hjet-own,gghphm-and-hw}.

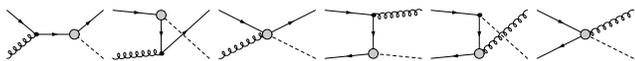
\begin{figure}[bt]
\begin{footnotesize}
\begin{picture}(378,40)(20,0)
\unitlength=1bp%
\begin{feynartspicture}(270,40)(6,1)
\FADiagram{}
\FAProp(0.,15.)(6.,10.)(0.,){/Straight}{1}
\FAProp(0.,5.)(6.,10.)(0.,){/Cycles}{0}
\FAProp(20.,15.)(14.,10.)(0.,){/Straight}{-1}
\FAProp(20.,5.)(14.,10.)(0.,){/ScalarDash}{0}
\FAProp(6.,10.)(14.,10.)(0.,){/Straight}{1}
\FAVert(6.,10.){0}
\FAVert(14.,10.){-1}

\FADiagram{}
\FAProp(0.,15.)(10.,14.)(0.,){/Straight}{1}
\FAProp(0.,5.)(10.,6.)(0.,){/Cycles}{0}
\FAProp(20.,15.)(10.,6.)(0.,){/Straight}{-1}
\FAProp(20.,5.)(10.,14.)(0.,){/ScalarDash}{0}
\FAProp(10.,14.)(10.,6.)(0.,){/Straight}{1}
\FAVert(10.,6.){0}
\FAVert(10.,14.){-1}

\FADiagram{}
\FAProp(0.,15.)(10.,10.)(0.,){/Straight}{1}
\FAProp(0.,5.)(10.,10.)(0.,){/Cycles}{0}
\FAProp(20.,15.)(10.,10.)(0.,){/Straight}{-1}
\FAProp(20.,5.)(10.,10.)(0.,){/ScalarDash}{0}
\FAVert(10.,10.){-1}

\FADiagram{}
\FAProp(0.,15.)(10.,14.)(0.,){/Straight}{1}
\FAProp(0.,5.)(10.,6.)(0.,){/Straight}{-1}
\FAProp(20.,15.)(10.,14.)(0.,){/Cycles}{0}
\FAProp(20.,5.)(10.,6.)(0.,){/ScalarDash}{0}
\FAProp(10.,14.)(10.,6.)(0.,){/Straight}{1}
\FAVert(10.,14.){0}
\FAVert(10.,6.){-1}

\FADiagram{}
\FAProp(0.,15.)(10.,14.)(0.,){/Straight}{1}
\FAProp(0.,5.)(10.,6.)(0.,){/Straight}{-1}
\FAProp(20.,15.)(10.,6.)(0.,){/Cycles}{0}
\FAProp(20.,5.)(10.,14.)(0.,){/ScalarDash}{0}
\FAProp(10.,14.)(10.,6.)(0.,){/Straight}{1}
\FAVert(10.,14.){0}
\FAVert(10.,6.){-1}

\FADiagram{}
\FAProp(0.,15.)(10.,10.)(0.,){/Straight}{1}
\FAProp(0.,5.)(10.,10.)(0.,){/Straight}{-1}
\FAProp(20.,15.)(10.,10.)(0.,){/Cycles}{0}
\FAProp(20.,5.)(10.,10.)(0.,){/ScalarDash}{0}
\FAVert(10.,10.){-1}

\end{feynartspicture}
\end{picture}
\end{footnotesize}

\caption{\label{MSSM-xtra-tops}
Additional topologies for the loop-induced 
process $q g \to q h^0$ and $q\bar q\to g h^0$ in the MSSM. The hatched circles
represent loops containing at least one gluino line. 
}
\end{figure}

The presence of superpartners in the loop contributions 
of the MSSM modifies the overall production rate for 
supersymmetric Higgs bosons compared to the production of 
SM Higgs bosons of equal mass. 
Moreover, there are new  Feynman graph topologies containing at least
one gluino line in the MSSM (see Fig.~\ref{MSSM-xtra-tops}) which also affect 
the angular distributions
and, at the level of hadronic processes,
change rapidity and transverse-momentum distributions
of the Higgs bosons or the jets, respectively.

\section{MSSM results}

In the following discussion we want to illustrate
the MSSM predictions for the
pseu\-do\-ra\-pi\-di\-ty, $\eta_\JET$, and transverse momentum ($p_T$) 
distributions of the
hadro\-nic processes $pp \to h^0 + \text{jet} + X$ 
and $p\bar p \to h^0 + \text{jet} + X$
and outline differences between MSSM and SM predictions.
To compare a given MSSM scenario 
with the SM, we choose the SM Higgs mass
to have the same value as the $h^0$ boson  
in that MSSM scenario.
For the numerical evaluation, we use the cuts 
\begin{align}
\label{the-cuts}
p_T & > 30\,\gev\,, & | \eta_{3} | & < 4.5\,,
\end{align}
which have been used in previous Standard Model
studies for the LHC~\cite{ADIKSS,Zmushko}.
Details of the calculation and the MSSM parameter constraints
taken into account 
can be found in \cite{hjet-own,hjet-own-distr}

We show here results for the $m_h^\MAX(400)$ scenario,
which is specified as follows.

The soft-breaking sfermion mass parameter is
set to $\MSUSY = 400\,\gev$.
The off-diagonal term $X_t$ ($= A_t-\mu\cot\beta$)
in the top-squark mass matrix is set to $2 \MSUSY$ ($=800\,\gev$)
The Higgsino and gaugino mass parameters have the settings
$\mu=- 200 \,\gev$,
$M_1 = M_2 = 200 \,\gev$,
$M_{\tilde g} = 800 \,\gev$.
When $\tb$ is changed, $A_t$ is changed accordingly to ensure $X_t = 2 \MSUSY$. 
The settings of the other soft-breaking scalar-quark Higgs couplings 
are $A_b = A_t$ and $A_f = 0$ ($f=e,\mu,\tau,u,d,c,s$).

In the $m_h^\MAX(400)$ scenario, small values of $m_A$ are still allowed.
Hence we examine two Higgs sector scenarios: $m_A = 110\,\gev$, $\tb=30$,
and $m_A = 400\,\gev$, $\tb=30$.
The former leads to the dominance of $b$-quark initiated processes,
while the latter is dominated by the 
loop-induced processes~\cite{hjet-own}.

\subsection{Differential cross sections at the LHC}

The crucial parameter determining the properties of $h^0$+jet production
in the MSSM is $m_A$ \cite{hjet-own}. 
For $m_A \lesssim 120\,\gev$ and $\tb$ not too small ($\gtrsim 5$) 
the $b$-quark initiated processes 
dominate the production rate by far, 
while for larger $m_A$ this role is taken over by the loop-induced
processes 
Accordingly, we split our discussion 
into the high-$m_A$ and low-$m_A$ cases.

\subsubsection{High {\boldmath{$m_A$}}}

The influence of 
rather light, yet not excluded, superpartners 
on the total hadronic cross section
has been demonstrated to be strong \cite{hjet-own}. 
In particular for the $m_h^\MAX(400)$ scenario with 
$\MSUSY=400\,\gev$,
the MSSM cross section for $m_A > 200\,\gev$ and any $\tb \in [1,50]$
is reduced by about $20 - 40\%$ compared to the SM.
Here, we investigate the impact on the shape of the
differential distributions
with respect to the SM.

The total hadronic cross section in the $m_h^\MAX(400)$ scenario 
is about 25\% smaller than in the SM. 
Yet, as far as the $\eta_\JET$ and $p_T$ dependent differences between
MSSM and SM are concerned, the same qualitative picture appears.
The variation of the relative difference $\delta$ with $\eta_\JET$ 
in the range $|\eta_\JET| < 4.5$ is about 2\% and 
with $p_T$ in the range $p_T \in [30\,\gev,1000\,\gev]$
is about 7\%.

\begin{figure}[htb]
{\setlength{\unitlength}{1cm}
\begin{picture}(6,5)
\put(0,0){\resizebox{.55\width}{.55\height}{
\includegraphics*{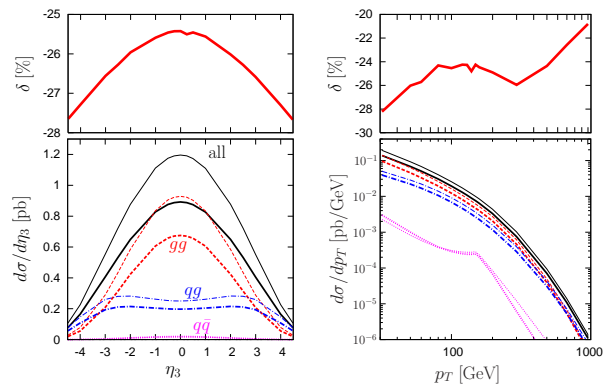}}}
\end{picture}
}
\caption{
\label{mh-max}
LHC, $m_{h}^\MAX(400)$ scenario with $\ma=400\,\gev$, $\tb=30$:
differential hadronic cross sections for Higgs + jet production and
the relative difference, $\delta$, between the MSSM and SM prediction.
Thick and thin lines correspond to the MSSM and SM prediction, respectively.
}
\end{figure}

\begin{figure}[h]
{\setlength{\unitlength}{1cm}
\begin{picture}(6,5.5)(0,0)
\put(1,.5){\resizebox{.5\width}{.5\height}{
\includegraphics*{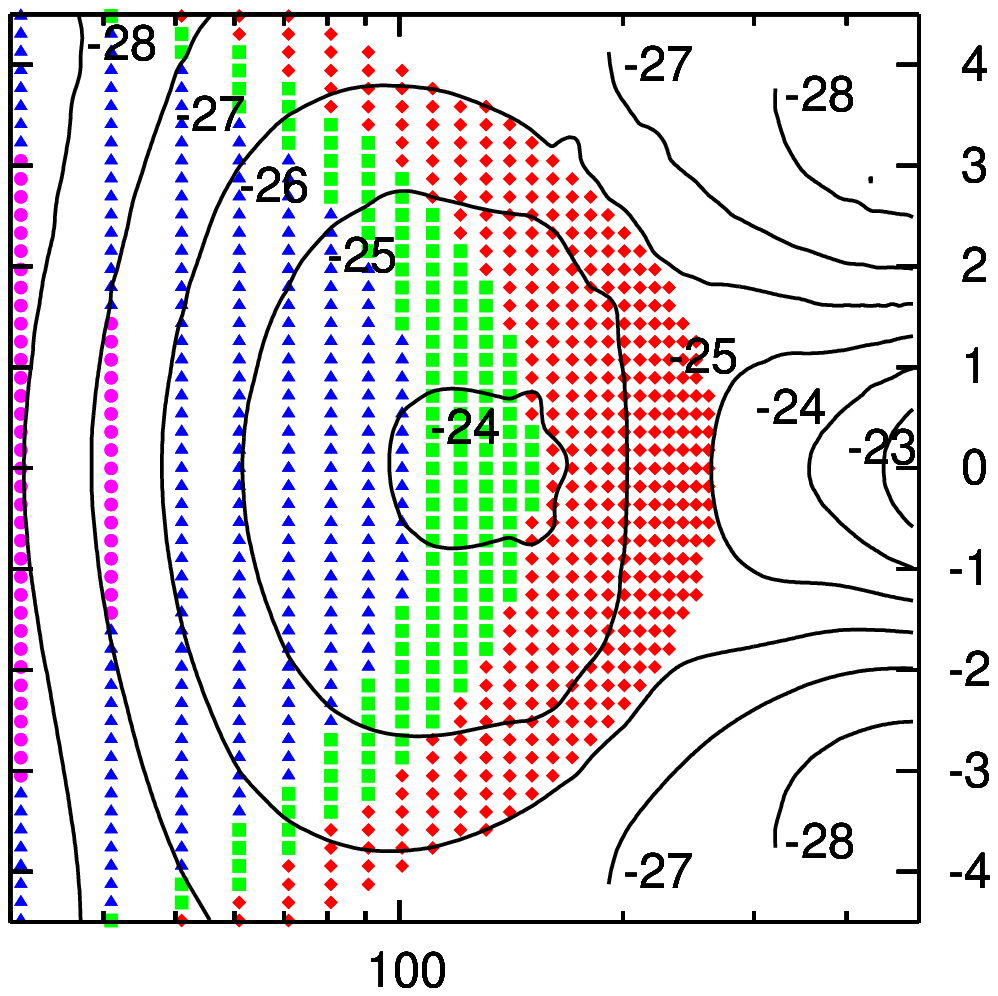}}}
\put(4,.3){\large $p_T\; [\gev]$}
\put(6.5,3){\large $\eta_\JET$}
\end{picture}
}
\caption{
\label{doubly-diff} Relative and absolute difference between
the MSSM and SM prediction for $d^2\sigma/d\eta_\JET d p_T$ at the LHC for 
the $m_{h}^\MAX(400)$ scenario with $\ma=400\,\gev$, $\tb=30$
as a function of $p_T$ and $\eta_\JET$.
Contour lines show the relative difference in \%, 
while 
diamonds ($\GNUPlotA{\scriptstyle\blacklozenge}$),
squares ($\GNUPlotB{\scriptstyle\blacksquare}$),
triangles ($\GNUPlotC{\scriptstyle\blacktriangle}$),
circles ($\GNUPlotD{\bullet}$),
refer to an absolute difference
in the range 
\GNUPlotA{0.1-0.5 $\fb/\gev$},
\GNUPlotB{0.5-1 $\fb/\gev$},
\GNUPlotC{1-5 $\fb/\gev$},
\GNUPlotD{5-10 $\fb/\gev$}
respectively.
In the white area the difference is less than $0.1\,\fb/\gev$.
}
\end{figure}

The relative difference between the MSSM and SM prediction
for the two-fold differential cross section $d^2\sigma/d p_T/d\eta_\JET$,
indicated by the contours in Fig.~\ref{doubly-diff}, 
shows a non-trivial behaviour
with an overall variation of more than 6\% in the depicted range,
$|\eta_\JET| < 4.5$ and $30\,\gev < p_T <500\,\gev$.
The differently shaped dots in Fig.~\ref{doubly-diff}
show the absolute difference between MSSM and SM, 
which gives an idea of the kinematical region where the LHC 
experiments may become sensitive to this difference.

Modifying the cuts may increase the sensitivity 
to deviations from the SM.
Guided by Fig.~\ref{doubly-diff}, we study 
the cross section $\sigma_f$ with rather soft forwardish jets
and $\sigma_c$ with harder more central jets:
\begin{align*}
\sigma_c & = \sigma\left(pp\to h^0 + j +X \right)|_{|\eta| < 1.5,\; p_T > 70\,\gev}
\,,\\
\sigma_f & = \sigma\left(pp\to h^0 + j +X \right)|_{
\stackrel{\scriptstyle 1.5 < |\eta| < 4.5,}{\scriptstyle 30\,\gev < p_T < 50\,\gev}}
\,.
\end{align*}
Table~\ref{thetable} shows the results,
where also the ratio
\begin{align}
\label{ratio}
R & = \frac{\sigma_c}{\sigma_f}\,,
\end{align}
and the relative difference between MSSM and SM
\begin{align}
\Delta & = \frac{R_\MSSM- R_\SM}{R_\SM}
\end{align}
are listed.
While each individual cross section in the MSSM and the SM
is still of the order of $1\,\pb$, 
which translates into $10^5$ raw events for an integrated luminosity 
of $100\,\fb^{-1}$, 
the MSSM ratio $R_\MSSM$ differs by $4.2\%$ compared to $R_\SM$.

\begin{table}[t]
\begin{center}
\begin{tabular}{c|cc}
quantity & SM & MSSM \\
\hline
$\sigma_c$
 & 1.448$\,\pb$ & 1.096$\,\pb$ \\
$\sigma_f$
 & 1.419$\,\pb$ & 1.031$\,\pb$  \\
\hline
$R = \sigma_c/\sigma_f$ 
 & 1.020 & 1.063 \\
\hline
$\Delta$ 
&\multicolumn{2}{c}{4.2\%}\\
\end{tabular}
\end{center}
\caption{\label{thetable} Cross section prediction
in the $m_h^\MAX(400)$ scenario
for Higgs + jet production with jets radiated into the central ($\sigma_c$) 
and forward part of the detector ($\sigma_f$), together with
their ratio $R$ and the relative difference between the MSSM and 
SM value for $R$, $\Delta$.
}
\end{table}

\subsubsection{Low {\boldmath{$m_A$}}}

\begin{figure}[hbt]
{\setlength{\unitlength}{1cm}
\begin{picture}(6,5)
\put(0,0){\resizebox{.55\width}{.55\height}{
\includegraphics*{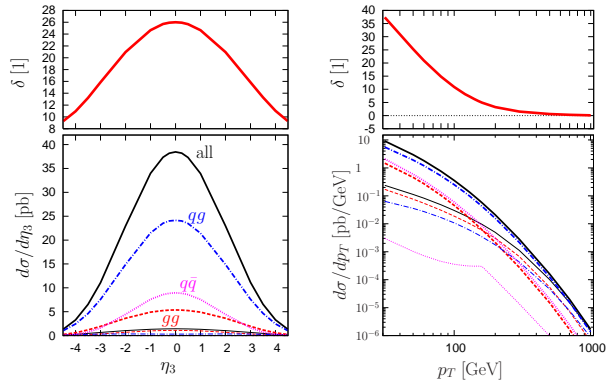}}}
\end{picture}
}
\caption{
\label{mh-max-smallma}
LHC, $m_{h}^\MAX(400)$ scenario with $\ma=110\,\gev$, $\tb=30$.
(See also caption of Fig.~\ref{mh-max}.)
}
\end{figure}

As an example for the low-$m_A$ case at the LHC we show results for 
the $m_h^\MAX(400)$ scenario in Fig.~\ref{mh-max-smallma}.
The change with respect to the SM is dramatic.
Due to the enhanced cross sections of the $b$-quark processes,
the quark-gluon scattering contribution dominates the cross section
and even the contribution from $q \bar q$ is larger than from gluon fusion.
The total hadronic cross section in the MSSM is 22 times higher 
than in the SM ($\approx 175\,\pb$).

Out of all jets allowed by our cuts~(\ref{the-cuts})
a larger fraction
of jets is radiated into the central part of the detector.
For instance, the fraction of jets 
produced with $|\eta_\JET| < 2$ compared 
to all jets allowed by the cuts 
is 93\% in the MSSM versus 85\% in the SM.
Correspondingly, the $p_T$ spectrum is much softer than in the SM,
yielding an enhanced rate for processes with jet transverse momenta 
below $850\,\gev$, e.g. by a factor of 10
for $p_T=100\,\gev$, and rates similar to the SM above $850\,\gev$.

\subsection{Differential cross sections at the Tevatron}

The typical hadronic cross section for Higgs + jet in the SM
expected at the Tevatron for the cuts $p_T > 30\,\gev$ and $|\eta_\JET| < 4.5$
is around $0.1\,\pb$ for Higgs masses around 100\,\gev, 
which is possibly not sufficient to be observable at the Tevatron.
Therefore, for the Tevatron only the MSSM scenarios with low
$m_A$ and $\tb$ not too small are of interest.
Those scenarios exhibit a cross section  enhanced by a factor of 
up to 30 compared to the SM \cite{hjet-own}.
This is due to the contribution of $b$-quark initiated processes
which become dominant because of the strongly enhanced Yukawa coupling 
of $b$-quarks to the Higgs boson $h^0$.

\begin{figure}[hbt]
{\setlength{\unitlength}{1cm}
\begin{picture}(6,5)
\put(0,0){\resizebox{.55\width}{.55\height}{
\includegraphics*{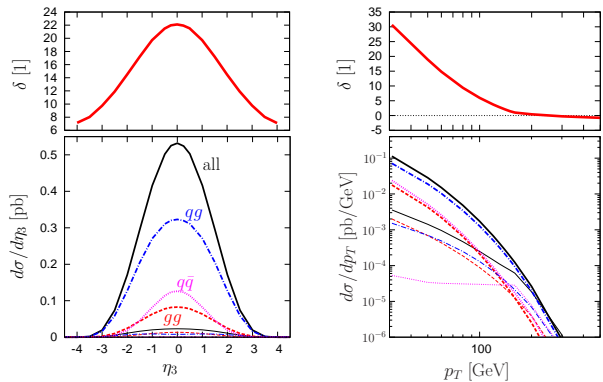}}}
\end{picture}
}
\caption{
\label{mh-max-smallma-tev}
Tevatron, $m_{h}^\MAX(400)$ scenario with $\ma=110\,\gev$, $\tb=30$.
(See also caption of Fig.~\ref{mh-max}.)
}
\end{figure}

Fig.~\ref{mh-max-smallma-tev} shows results for 
the same low-$m_A$ scenario as just described for the LHC in the 
previous paragraph.
Very similar to the LHC case, we see a strongly enhanced total
hadronic cross section with a softer $p_T$ spectrum and a larger 
fraction of jets radiated into the central part of the detector than in the SM.

\section{Summary}

We have calculated pseudorapidity and transverse momentum 
distributions
for the MSSM $h^0$ + high-$p_T$ jet production cross section
at the LHC and the Tevatron.
For scenarios with large $m_A$,
the loop-induced processes dominate the cross section, and
superpartners can have a significant impact when they are not too heavy.
For small $m_A$, the Yukawa couplings of the $b$-quarks are enhanced and 
hence the cross section is dominated by $b$-quark induced 
tree-level parton reactions.
The example investigated here,
the $m_h^\MAX(400)$ scenario, 
shows a strongly enhanced hadronic cross section compared to the SM,
by a factor of more than 20.
Such a scenario predicts for both LHC and Tevatron
a softer $p_T$ spectrum, with a fraction
of jets radiated into the central part of the detector
larger than in the SM.


\begin{thebibliography}{999}

\bibitem{CMS-TDR}
CMS Physics Technical Design Report, Vol. II, CERN/LHCC 2006-021.


\bibitem{hjet-sm}
R.~K.~Ellis {\it et al.},
Nucl.\ Phys.\ B {\bf 297}, 221 (1988);
U.~Baur and E.~W.~Glover,
Nucl.\ Phys.\ B {\bf 339}, 38 (1990);
M.~Chaichian {\it et al.},
Phys.\ Lett.\ B {\bf 198}, 416 (1987)
[Erratum-ibid.\ B {\bf 205}, 595 (1987).



















\bibitem{ADIKSS}
S.~Abdullin, M.~Dubinin, V.~Ilyin, D.~Kovalenko, V.~Savrin and N.~Stepanov,
Phys.\ Lett.\ B {\bf 431} (1998) 410.

\bibitem{Zmushko}
V.~V.~Zmushko, ATL-PHYS-2002-020, IHEP-2002-23.

\bibitem{mellado-etal}
B.~Mellado, W.~Quayle and S.~L.~Wu,
Phys.\ Lett.\ B {\bf 611} (2005) 60.

\bibitem{new-mellado-etal}
B.~Mellado, W.~Quayle and S.~L.~Wu,
  arXiv:0708.2507 [hep-ph], to appear in Phys.\ Rev.\  D.

\bibitem{AMP}
C.~Anastasiou, K.~Melnikov and F.~Petriello,
Phys.\ Rev.\ Lett.\  {\bf 93} (2004) 262002,
Nucl.\ Phys.\ B {\bf 724} (2005) 197.


\bibitem{grazzini-etal}
G.~Bozzi, S.~Catani, D.~de Florian and M.~Grazzini,
Phys.\ Lett.\ B {\bf 564} (2003) 65;
S.~Catani, D.~de Florian, M.~Grazzini and P.~Nason,
JHEP {\bf 0307} (2003) 028.

\bibitem{kunszt-etal}
D.~de Florian, M.~Grazzini and Z.~Kunszt,
Phys.\ Rev.\ Lett.\  {\bf 82} (1999) 5209.

\bibitem{kulesza-etal}
D.~de Florian, A.~Kulesza and W.~Vogelsang,
JHEP {\bf 0602} (2006) 047.

\bibitem{bghb-QCD}
J.~Campbell, R.~K.~Ellis, F.~Maltoni and S.~Willenbrock,
  Phys.\ Rev.\  D {\bf 67} (2003) 095002;
S.~Dawson, C.~B.~Jackson, L.~Reina and D.~Wackeroth,
  Phys.\ Rev.\ Lett.\  {\bf 94} (2005) 031802;
S.~Dittmaier, M.~Kramer and M.~Spira,
  Phys.\ Rev.\  D {\bf 70} (2004) 074010.

\bibitem{hjet-own}
O.~Brein and W.~Hollik,
Phys.\ Rev.\ D {\bf 68} (2003) 095006.

\bibitem{BField-etal}
B.~Field, S.~Dawson and J.~Smith,
Phys.\ Rev.\ D {\bf 69} (2004) 074013.

\bibitem{langenegger-etal}
U.~Langenegger, M.~Spira, A.~Starodumov and P.~Trueb,
hep-ph/0604156.

\bibitem{dawson-jackson}
S.~Dawson and C.~B.~Jackson,
  arXiv:0709.4519 [hep-ph].

\bibitem{gghphm-and-hw}
O.~Brein and W.~Hollik,
Eur.\ Phys.\ J.\ C {\bf 13} (2000) 175;
O.~Brein, W.~Hollik and S.~Kanemura,
Phys.\ Rev.\ D {\bf 63} (2001) 095001.


\bibitem{hjet-own-distr}
O.~Brein and W.~Hollik,
  Phys.\ Rev.\  D {\bf 76} (2007) 035002.






\end{thebibliography}
 \end{document}